\documentclass[groupedaddress,nofootinbib,showpacs,showkeys,amssymb,eqsecnum,aps]{revtex4-2}
\usepackage[pdftex]{graphicx}
\usepackage{amsmath} 
\usepackage{epsf}
\usepackage{color}
\usepackage{verbatim}                                                                         
\usepackage[pdftex]{hyperref}
\newcommand{\be}{\begin{equation}}                                                                   
\newcommand{\ee}{\end{equation}}                                                                         

\begin{document}

\title{Quantum gravity in a general background gauge}
\author{F. T. Brandt}
\email{fbrandt@usp.br}
\affiliation{Instituto de F\'{\i}sica, Universidade de S\~ao Paulo, S\~ao Paulo, SP 05508-090, Brazil}
\author{J. Frenkel}
\email{josiffrenkel@gmail.com}
\affiliation{Instituto de F\'{\i}sica, Universidade de S\~ao Paulo, S\~ao Paulo, SP 05508-090, Brazil}
\author{D. G. C. McKeon}
\email{dgmckeo2@uwo.ca}
\affiliation{
Department of Applied Mathematics, The University of Western Ontario, London, ON N6A 5B7, Canada}
\affiliation{Department of Mathematics and Computer Science, Algoma University,
Sault St.Marie, ON P6A 2G4, Canada}
\date{\today}
\begin{abstract}
We deduce, in a general background gauge, the counter-term Lagrangian for pure quantum  gravity to one-loop order. As an application, we evaluate the leading quantum correction to
the classical gravitational potential, generated by the vacuum polarization. We find that, in specific background gauges, this yields the complete result for the one-loop quantum corrections to the Newtonian potential. This approach is also applied to calculate the $\ln(T)$ contributions in quantum gravity at high-temperature.
\end{abstract}                                                                       
\maketitle                     

\section{Introduction}
Einstein's general relativity, as defined by the Einstein-Hilbert action, may be regarded as a non-renormalizable effective field theory, which is 
expected to arise in the low-energy limit of some fundamental quantum theory. This theory appears 
to adequately describe the gravitational interactions which occur at low energies .  Much work has 
already been done on quantum gravity treated as an effective field theory \cite{F1,Donoghue:1993eb,F2,F3}. Using the Feynman rules 
deduced from the Einstein-Hilbert Lagrangian,  one can perform in quantum gravity perturbative  calculations of loop diagrams \cite{F4,Capper:1978yf,Kallosh:1978wt,F5,Brandt:2016eaj,F6,F7,Kalmykov:1998cv,F7a,Bastianelli:2022pqq,Toms:2011zza}. The corresponding contributions require an infinite number of counter-terms allowed by the gauge symmetry,  to cancel out all the ultraviolet divergences in this 
effective field theory.

In this context, the background field method \cite{F8,F9,Grisaru:1975ei,F10,F11,F12,F13,F14} has been much employed in the calculation of radiative effects in quantum gravity since this procedure preserves the gauge invariance of the background field.  It has been first shown by 't~ Hooft and Veltman \cite{F4} that on mass-shell,  pure gravity is renormalizable to one-loop order.  This analysis has been done in a particular  background gauge.  

The purpose of the present work is to examine such calculations in a general background  gauge characterized by a gauge parameter $\xi$.  
We thus consider  the behavior of the graviton self-energy in any space-time
\hbox{dimension $d$ \cite{tHooft:1972tczBollini:1972ui}}, which turns
out  to explicitly depend on the gauge parameter $\xi$. Due to the
Ward identities which hold in the background gauge,  this behavior implies that all higher-point
Green's functions will depend as well on this gauge parameter.  These Ward identities reflect the
gauge invariance of the theory under background gauge transformations. In this way,  we infer that 
for pure gravity in $d = 4 - 2\epsilon$  space-time dimensions, the counter-term Lagrangian in a general background gauge may be written to one-loop order in the form 
\be\label{eq1}
   {\cal L}_{one-loop}^{div}=\frac{\sqrt{-\bar g}}{8\pi^2(4-d)}\left\{
   \left[
\frac{1}{120} + \frac{1}{6}(\xi-1)^2
     \right] \bar R^2
   +\left[
\frac{7}{20} + \frac{\xi(\xi-1)}{3}     \right]\bar R^{\mu\nu} \bar R_{\mu\nu}     
   \right\}
\ee
which reduces to the result obtained by 't Hooft and Veltman \cite{F4} for $\xi=1$.  Here, $\bar R_{\mu\nu}$ denotes the Ricci tensor, $\bar R =\bar R^{\mu\nu} \bar g_{\mu\nu}$  is the curvature scalar and $\bar g_{\mu\nu}$  is the background metric.

In spite of the lack of predictivity of the theory at high energy (short distance), it has been shown that the leading low-energy (long distance) quantum corrections can be consistently evaluated in quantum gravity \cite{Donoghue:1993eb,F2,F3}.  Such contributions,  that are due to the interactions of massless particles at low energy,  include non-analytic terms of the form $\ln(-k^2)$,  which in the low energy limit yield the leading quantum corrections to the classical Newtonian potential ($k$ is the momentum transfer involved in the process). In this work,  we evaluate the corresponding contribution due to the self-energy of the graviton,  which turns out to be gauge -dependent.  However,  when one considers the quantum corrections arising from all diagrams that contribute to the physical gravitational potential, the gauge-dependent terms  should cancel out. 

The outline of the paper is as follows. In Sec. \ref{sec2} we discuss the quantization of general relativity in a general background gauge, following the work of  't Hooft and Veltman \cite{F4}.
In Sec. \ref{sec3}, we examine in this gauge the graviton self-energy to one loop order. As an application, we evaluate the corresponding quantum correction to the classical gravitational potential due to two heavy masses. We find that this contribution is a gauge-dependent quantity.
We show that in the background gauges $\xi=(2\pm\sqrt{13})/3$,
the complete result for such quantum corrections arises just from the vacuum polarization.
In Sec. \ref{sec4}, we present the Ward identities and  deduce the gauge invariant counter-term  Lagrangian  to one loop order. We conclude the paper with a summary of the results in Sec \ref{sec5},
where we also briefly discuss the $\ln(T)$ contributions of the graviton amplitudes at high temperature $T$. The Feynman rules are derived in Appendix \ref{appA} and  several details of the calculation of the graviton self-energy are  given in Appendix \ref{appB}.
In Appendix \ref{appC}, we consider the consequences of employing a Lagrangian multiplier field to restrict radiative corrections to one loop order.

\section{Quantization with a general gauge-fixing Lagrangian}\label{sec2}
In order to quantize the theory of general relativity, we start from the Einstein-Hilbert Lagrangian 
\be\label{eq2}
{\cal L}_g = \sqrt{-g} \frac{2}{\kappa ^2} R,
\ee
where $\kappa^2 = 32\pi G$ and $G$ is Newton's constant. The metric tensor $g_{\mu\nu}$ is written as
\be\label{eq3}
g_{\mu\nu} = \bar g_{\mu\nu} + \kappa h_{\mu\nu},
\ee
where $\bar g_{\mu\nu}$ is the background field which is assumed to approach the classical vacuum at infinity (being arbitrary elsewhere) and $h_{\mu\nu}$ is the quantum field.

Expanding the Lagrangian \eqref{eq2} in the quantum field, one obtains the  following quadratic Lagrangian
\begin{eqnarray}\label{eq4}
 {\cal L}^{(2)}_g & = & \sqrt{-\bar g}\Biggl[{1 \over 2} \bar D_{\alpha} h_{\mu \nu} \bar D^{\alpha}
h^{\mu \nu} - {1 \over 2} \bar D_{\alpha} h \bar D^{\alpha} h + \bar D_{\alpha} h
\bar D_{\beta} h^{\alpha \beta}
-\bar D_{\alpha} h_{\mu \beta} \bar D^{\beta} h^{\mu \alpha}
\nonumber \\ 
&+&  \bar{R} \left(
{1 \over 4} h^2 - {1 \over 2} h_{\mu \nu}h^{\mu \nu} \right)
+  \bar{R}^{\mu \nu} \left(2 h^{\alpha} _{~\mu} h_{\nu \alpha}
- h h_{\mu \nu} \right) \Biggr]. 
\end{eqnarray}
where  $h=h^\lambda_\lambda$ and $\bar D_\alpha$ is the covariant derivative with respect to the quantum field.
In order to quantize the above quadratic Lagrangian one has to fix the gauge of the quantum 
field in a way which preserves the gauge invariance under background field transformations. This can be achieved by introducing the gauge-fixing Lagrangian 
\begin{equation}\label{eq5}
{\cal L}_{gf} = \frac{1}{\xi}\sqrt{-\bar g}\left[\left( \bar D^{\nu} h_{\mu \nu} - {1 \over
  2} \bar D_{\mu} h \right) \left( \bar D_{\sigma}
h^{\mu\sigma} - {1 \over 2} \bar D^{\mu} h \right)\right]
\end{equation}
where $\xi$ is a generic gauge parameter. When $\xi=1$, the above expression reduces to the background harmonic
gauge used in \cite{F4}. After some work, the corresponding ghost Lagrangian is found to be
\begin{equation}\label{eq6}
{\cal L}_{gh} = \sqrt{-\bar g}\; c^{* \mu} \left[ \bar D_{\lambda} \bar D^{\lambda}
\bar{g}_{\mu \nu} - \bar R_{\mu \nu} \right] c^{\nu},
\end{equation}
where the ghost fields ${c^\star}^\mu$ and $c^\nu$ are fermionic vector fields.
We note here that the actions corresponding to the \eqref{eq4}, \eqref{eq5} and \eqref{eq6},
are separately invariant under the  gauge transformations
\begin{subequations}
\be\label{eq7}
\delta \bar g_{\mu\nu} = \omega^\gamma \partial_\gamma \bar g_{\mu\nu}
+\bar g_{\mu\gamma} \partial_\nu \omega^\gamma +\bar g_{\nu\gamma} \partial_\mu \omega^\gamma 
= \bar D_\mu \omega_\nu+\bar D_\nu \omega_\mu ,
\ee
and
\be\label{eq8}
\delta h_{\mu\nu} = \omega^\gamma \partial_\gamma h_{\mu\nu}
+h_{\mu\gamma} \partial_\nu \omega^\gamma + h_{\nu\gamma} \partial_\mu \omega^\gamma ,
\ee
\end{subequations}
where $\omega^\gamma$ is an infinitesimal parameter.
The total action can then be used to define the propagator of the quantum field and extract the corresponding Feynman rules, as shown in the Appendix \ref{appA}. The graviton propagator is given by
\begin{eqnarray}\label{eq9}
{\cal D}^{grav}_{\mu\nu\,\alpha\beta}(p) &=&\frac{i}{p^2+i\epsilon}\Biggl[
 \frac{1}{2}  \left(\eta_{\alpha \nu } \eta_{\beta \mu }
+  \eta_{\alpha \mu } \eta_{\beta \nu }
-\frac{2 \eta_{\alpha \beta } \eta_{\mu \nu}}{d-2}\right)
\nonumber \\
&+&\frac{\xi -1}{2 p^2} \left({ p_{\beta } p_{\nu } \eta_{\alpha\mu }}
+{ p_{\beta } p_{\mu } \eta_{\alpha\nu }}
+{ p_{\alpha } p_{\nu } \eta_{\beta\mu }}
+{ p_{\alpha } p_{\mu } \eta_{\beta\nu }}\right)\Biggl].
\end{eqnarray}
Owing to the complexity of the Feynman rules in a general background gauge, we do not quote here
these rules, which were derived entirely by computer.

\section{The one-loop graviton self-energy}\label{sec3}
\begin{figure}[b!]
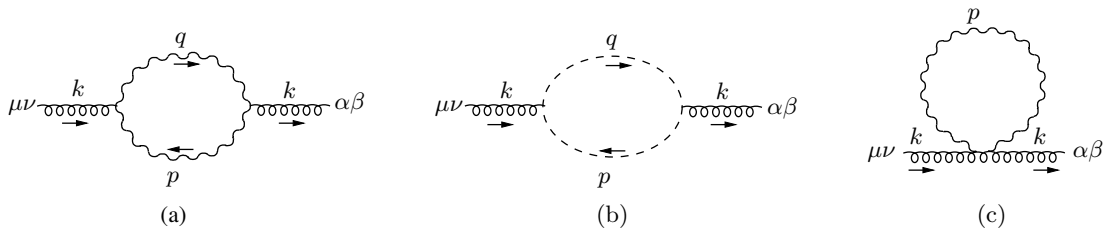

\input{graviton_loopN.pspdftex}\qquad \qquad
\input{ghost_loop.pspdftex}\qquad \qquad
\input{tad_pole.pspdftex}
\caption{One-loop contributions to $\langle \bar h\bar h \rangle$. The curly, wavy and dashed lines are associated with the background fields, the quantum fields and the ghost fields respectively. The arrows indicate the direction of momenta and $q=k+p$. Diagram (c) as well as a similar diagram with a ghost loop, do not contribute when we employ dimensional regularization.}\label{fig1}
\end{figure}

The Feynman diagrams contributing at one loop to the graviton self-energy are shown in Fig. (\ref{fig1}). The divergent contributions of these diagrams to the graviton self-energy are
evaluated in any space-time dimensions $d$ in the Appendix \ref{appB}. As shown in Eq. \eqref{PiLL},
for $d=4-2 \epsilon$, the divergent terms are given by 
%
\be\label{eq10}
\Pi^{div}_{\mu\nu ,\,\alpha\beta}(k)  = \frac{\kappa^2}{16\pi^2}\frac{\mu^{4-d}}{4-d} k^4\left\{
4 \, c_1(\xi) \, L_{\mu\nu} L_{\alpha\beta} + c_2(\xi) \left[
L_{\mu\nu} L_{\alpha\beta} + \frac{1}{2} \left(L_{\alpha\mu} L_{\beta\nu} + L_{\alpha\nu} L_{\beta\mu}\right) 
\right]
\right\},
\ee
where $L_{\mu\nu} = \frac{k_\mu k_\nu}{k^2} - \eta_{\mu\nu}$ and $c_1(\xi)$, $c_2(\xi)$ are gauge-dependent constants given by
\be\label{eq11}
c_1(\xi) = \left[\frac{1}{120}+\frac{(\xi-1)^2}{6}\right] ; \;\;\;\; c_2(\xi) = \left[\frac{7}{20}+\frac{\xi(\xi-1)}{3}\right].
\ee
We note here that Eq. \eqref{eq10} is transverse with respect to the momentum $k$. As we will show in the next section, this behavior is a consequence of the Ward identities which hold in the background gauge.

In Eq. \eqref{eq10} an arbitrary scale factor $\mu$, with dimensions of mass, has been inserted on dimensional grounds. Since $k$ is the only other dimensional quantity, one can see that
\be\label{eq12}
\frac{1}{4-d}\left(\frac{\mu^2}{-k^2}\right)^{(4-d)/2}\approx \frac{1}{4-d} - \frac 1 2\ln\left(\frac{-k^2}{\mu^2}\right)+ {\cal O}(4-d)
\ee
This relation allows us to extract directly from Eq. \eqref{eq10},
the non-analytic $\ln(-k^2)$ contribution
\be\label{eq13}
\Pi^{\ln(-k^2)}_{\mu\nu ,\,\alpha\beta}  = -\frac{G}{\pi} \,\ln(-k^2)\,k^4\left\{
4 \, c_1(\xi) \, L_{\mu\nu} L_{\alpha\beta} + c_2(\xi) \left[
L_{\mu\nu} L_{\alpha\beta} + \frac{1}{2} \left(L_{\alpha\mu} L_{\beta\nu} + L_{\alpha\nu} L_{\beta\mu}\right) 
\right]
\right\},
\ee
where we have used the fact that $\kappa^2 = 32\pi G$.
In order to calculate the quantum corrections to the gravitational potential, we employ the following
coupling of the external background field to the energy momentum tensor $T^{\mu\nu}$ of the matter fields
\be\label{eq14}
{\cal L}_I = -\frac{\kappa}{2} \bar h_{\mu\nu} T^{\mu\nu},
\ee
where we have defined $\bar g_{\mu\nu} = \eta_{\mu\nu} + \kappa \bar h_{\mu\nu}$.
For external spineless sources described by the Lagrangian
\be\label{eq15}
{\cal L}_{M} =\frac{\sqrt{-g}}{2}\left(g^{\mu\nu}\partial_\mu\phi\partial_\nu\phi - m^2\phi^2\right),
\ee
the tensor is
\be\label{eq16}
T_{\mu\nu} = \partial_\mu\partial_\nu\phi - \frac 1 2\eta_{\mu\nu}(\partial_\lambda\phi\partial^\lambda\phi-m^2\phi^2).
\ee
Using this result in Eq. \eqref{eq14}, we obtain in momentum space a graviton matter coupling of the form
\be\label{eq17}
V_{\mu\nu}(p)=-\frac{\kappa}{2}\left[p_\mu p^\prime_\nu + p^\prime_\mu p_\nu - \eta_{\mu\nu} (p\cdot p^\prime-m^2)\right]
\ee
One can verify that this vertex is transverse with respect to $k$ when the scalar particles are on-shell.

\begin{figure}[t!]
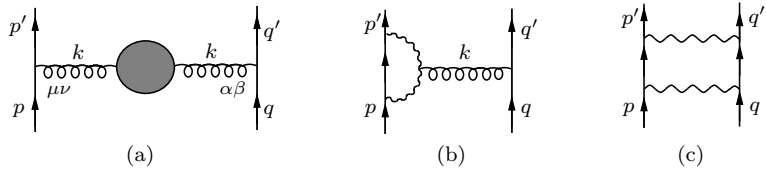

\input potentialA.pspdftex\qquad \qquad
\input potentialB.pspdftex\qquad \qquad
\input potentialC.pspdftex
\caption{Examples of Feynman diagrams which yield corrections to the gravitational potential.}
\label{fig2}
\end{figure}

We are now in a position to calculate the correction to the gravitational potential coming from the
Feynman diagram shown in Fig. (\ref{fig2}-a), where the blob denotes the graviton self-energy.
To this end, we will also need to define a propagator for the background field. The most general 
expression for this propagator is similar to that given in Eq. \eqref{eq9}, with $\xi$ replaced by
$\bar\xi$.
However, since a momentum  $k$  contracted with the self-energy or the vertex function gives a vanishing result,
such a propagator may be effectively replaced, in $4$ dimensions, by the de Donder propagator
\be
\bar {\cal D}_{\mu\nu\,\alpha\beta} =\frac{i}{k^2+i\epsilon}
 \frac{1}{2}  \left(\eta_{\alpha \nu } \eta_{\beta \mu }+  \eta_{\alpha \mu } \eta_{\beta \nu }
- {\eta_{\alpha \beta } \eta_{\mu \nu}}\right)
\ee
which is independent of the gauge parameter $\bar\xi$.
We can see from the figure (\ref{fig2}-a) that we will need to evaluate the quantity
$\bar {\cal D}^{\mu\nu, \rho\sigma}\Pi_{\rho\sigma,\,\lambda\delta} \bar{\cal D}^{\lambda\delta, \alpha\beta}$
where $\Pi_{\rho\sigma,\,\lambda\delta}$ can be obtained from Eq. \eqref{eq13}. Here, all terms involving         $k_\rho$, $k_\lambda$ etc, can be dropped
because $k$ contracted with the vertex gives a vanishing result. This yields the expression 
(see Eq. \eqref{B18a})
\begin{eqnarray}\label{eq19}
\Delta \tilde V_{(2a)}(k) &=&\frac{V_{\mu\nu}(p)}{2 p_0}
\bar {\cal D}^{\mu\nu, \rho\sigma}\Pi_{\rho\sigma,\,\lambda\delta}\bar{\cal D}^{\lambda\delta, \alpha\beta}  
\frac{V_{\alpha\beta}(q)}{2 q_0}
\nonumber \\ &=&
\frac{G}{\pi}\frac{\ln(-k^2)}{4 p_0 q_0}
V_{\mu\nu}(p)
\left[
c_1(\xi) \eta^{\mu\nu} \eta^{\alpha\beta} + c_2(\xi)\frac{\eta^{\alpha\mu}\eta^{\beta\nu}+\eta^{\alpha\nu} \eta^{\beta\mu}}{2}
 \right]V_{\alpha\beta}(q),
\end{eqnarray}
where the factors in the denominators account for the normalization of states.
We will evaluate this quantity in the case involving two heavy particles with mass $m$ , by taking the non-relativistic static limit $p\approx (m,0)$ in Eq. \eqref{eq19}. We then get  
\be\label{eq20}
\Delta\tilde V_{(2a)}(k) \approx G^2 m^2\ln({\vec k}^2)\left[\frac{43}{15}+\frac{4}{3}(\xi-1)(3\xi-1)\right],
\ee
where we used Eq. \eqref{eq17} and the constants $c_1(\xi)$ and $c_2(\xi)$ given in Eq. \eqref{eq11}.
This can be transformed to coordinate space by Fourier transforming, by using that 
\be\label{eq21}
\int\frac{d^3 k}{(2\pi)^3} e^{-i\vec k\cdot\vec r} \ln({\vec k}^2) = -\frac{1}{2\pi}\frac{1}{r^3}
\ee
We thus obtain for the correction generated by the graviton self-energy, the result 
(reinstating factors of $\hbar$ and $c$)
\be\label{eq22}
\Delta V_{(2a)}(r) = -\left[\frac{43}{30} + \frac{2}{3} (\xi-1)(3\xi-1)\right]
\frac{G\hbar}{\pi c^3 r^2}\frac{ G m^2}{r}.
\ee
This  expression reduces, for $\xi=1$, to the corresponding result obtained in Ref. \cite{F15}. Thus, we see that in a general background gauge, individual  corrections to the Newtonian potential may be gauge dependent. There are many other diagrams that can yield corrections to this potential, as shown for 
example in Figs. (\ref{fig2}-b) and (\ref{fig2}-c) \cite{F15}. Since the gravitational  potential is a physical quantity, the gauge-dependent terms should cancel out when adding all contributions. As shown in Refs. \cite{F15,F16}, the total result  obtained in the gauge $\xi=1$, is given in this case by
\be\label{eq23}
\Delta V(r) = -\frac{41}{10}
\frac{G\hbar}{\pi c^3 r^2}\frac{ G m^2}{r}.
\ee
By matching the results given in Eqs. \eqref{eq22} and \eqref{eq23}, one finds
that in the gauges $\xi=(2\pm\sqrt{13})/3$,
the complete  result would arise just from the correction generated by the graviton self-energy.

\section{The counter-term Lagrangian at one-loop}\label{sec4}
In the background field method, the gauge invariance of the effective action may be expressed as
\begin{eqnarray}\label{eq24}
  \delta \bar g_{\mu\nu} \frac{\delta\Gamma}{\delta \bar g_{\mu\nu}} = - 2\omega_\nu\bar D_\mu
\frac{\delta\Gamma}{\delta \bar h_{\mu\nu}} = 0,
\end{eqnarray}
where we have used the Eq. \eqref{eq7} and wrote the background field
in the form $\bar g_{\mu\nu}=\eta_{\mu\nu}+\kappa\bar h_{\mu\nu}$.
Taking the functional derivative of Eq. \eqref{eq24}
with respect to $\bar h_{\alpha\beta}$, evaluated at $\bar h=0$, one gets
\be\label{eq25}
\partial^\mu \Pi_{\mu\nu ,\alpha\beta} = 0
\ee
This  condition requires  that the graviton self-energy should be a transverse function, a property
which is explicitly shown in Eq. \eqref{PiLL}.
By taking the functional derivative of Eq. \eqref{eq24} with respect to $\bar h_{\alpha\beta}$ and $\bar h_{\rho\sigma}$ , evaluated at $\bar h = 0$,  we obtain the equation 
\be\label{eq26}
\partial^\mu V_{\mu\nu ,\alpha\beta ,\rho\sigma} = \frac{\kappa}{2}\left[ 
\eta_{\nu\alpha}\partial^\mu\Pi_{\mu\beta ,\rho\sigma}+\eta_{\nu\beta}\partial^\mu\Pi_{\mu\alpha ,\rho\sigma}
-\partial_\nu\Pi_{\alpha\beta ,\rho\sigma}
\right],
\ee
which relates the 3-point function to the 2-point function. Such Ward identities, which reflect the
gauge invariance of the effective action, may similarly be obtained for higher point functions (see Eqs. \eqref{Gward1}) 

This property of the effective action  implies that the effective counter-term Lagrangian must
be invariant, so it may be expressed in terms of invariant functions under background gauge transformations. The only invariant terms with four derivatives are
$\sqrt{-\bar g} \bar R_{\alpha\beta\mu\nu} \bar R^{\alpha\beta\mu\nu}$, $\sqrt{-\bar g} \bar R_{\mu\nu} \bar R^{\mu\nu}$ and $\sqrt{-\bar g} \bar R^2$.
But owing to a special relation  which holds in 4-dimensions among  these terms
(see Eq. \eqref{GB0}), one may express the effective one-loop counter-term Lagrangian in terms of just two such invariants, as
\be\label{eq27}
         {\cal L}^{(div)}_{1\;\, loop} = \frac{\sqrt{-\bar g}}{8\pi^2(4-d)}\left[c_1(\xi)\bar R^2
         + c_2(\xi)\bar R_{\mu\nu}  \bar R^{\mu\nu}\right],
\ee
where $c_1(\xi)$ and $c_2(\xi)$ are some coefficients. These may be fixed by comparing the result of the 2-point 
function obtained from \eqref{eq27}  with that calculated for the graviton self-energy
in Eqs. \eqref{eq10} and \eqref{eq11}. In this way, we obtain in pure  gravity the counter-term Lagrangian given in Eq. \eqref{eq1}. The fact that the 
coefficients in Eq. \eqref{eq27} are gauge-dependent
is not surprising since such structures vanish on mass-shell and can be absorbed by a field renormalization, which is not an observable quantity.

If we add matter fields , as we did in Sec. \ref{sec3}, these structures will no longer vanish on shell. But the gauge dependent parts of the Green functions with external background fields will remain unchanged. The effect of the matter fields is to add extra gauge-invariant  contributions to the Lagrangian. For example, introducing scalar fields as done
in Sec. \ref{sec3},  would yield  corrections arising from internal matter loops of the form \cite{F4}
\be\label{eq28}
{\cal L}^{(div)}_{M}  =  \frac{\sqrt{-\bar g}}{8\pi^2 (4-d)}\left[\frac{1}{240}\bar R^2+\frac{1}{120}\bar R_{\mu\nu} \bar R^{\mu\nu}\right].
\ee
The above features explain the gauge-dependence of the individual amplitudes computed in this framework.


\section{Discussion}\label{sec5}

We examined some features of quantum gravity, in a general background gauge to one loop order. In
this context, we derived an extension of the counter-term Lagrangian obtained by 't Hooft and Veltman,  
which is characterized by a gauge dependence of the coefficients that occur in this Lagrangian.
We studied the graviton self-energy and verified, in Appendix \ref{appB}, that the contributions arising from graviton and ghosts loops are separately transverse. This feature is a consequence of the fact that, in the background field method, the ghost Lagrangian is by itself invariant under background gauge transformations. 

We applied these results to the calculation of the quantum correction to the classical gravitational
potential generated by the graviton self-energy, which is a gauge-dependent function.
Since the Newtonian potential is a physical quantity, the sum of all quantum corrections
should be gauge-independent. The complete result obtained 
in the background harmonic gauge $\xi=1$ emerges from a summation of many Feynman diagrams \cite{F15,F16}.
We found that in the background gauges $\xi=(2\pm\sqrt{13})/3$, the 
full result arises just from the quantum correction generated by the vacuum polarization. 
Such a quantum correction is exceedingly small, being about $10^{-48}$ at $r=10^{-10} \, m$.

\begin{figure}[h!]
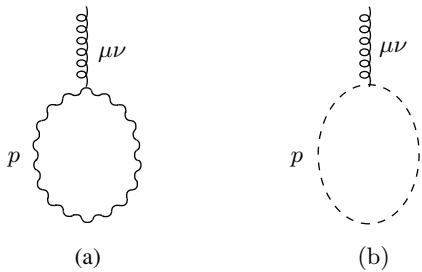

\input GravTad.pspdftex\qquad\qquad\qquad
\input GhostTad.pspdftex
\caption{Diagrams contributing to the thermal one-point graviton function.}
\label{fig3}
\end{figure}

Another useful application of the present approach concerns the graviton amplitudes at finite \hbox{temperature~$T$}.  These are of interest in quantum gravity both in their own right as well as for their
potential cosmological applications. Such amplitudes have a leading $T^4$ behavior at high-temperatures \cite{Brandt:1991qn,Brandt:1998hd}. For example, the one-point thermal graviton function shown in 
Fig. \ref{fig3}, which is related to the thermal energy-momentum tensor, is given by the expression
\begin{eqnarray}\label{eq51}
  \Gamma^{\mu\nu}_{thermal}
&=&\frac{\kappa \pi^2 T^4}{90}\left(\eta^{\mu\nu} - 4\,\eta^{\mu 0}\eta^{\nu 0}\right)
= \left.\frac{\delta S^{static}[\bar g]}{\delta\bar g_{\mu\nu}}\right|_{\bar g_{\mu\nu} = \eta_{\mu\nu}} 
\;\;\;\; (\mbox{for } d=4),
\end{eqnarray}
where $S^{static}$ is the static limit of the leading high-temperature thermal effective action given by
\cite{Brandt:2009ht} 
\be
S^{static}[\bar g] = \frac{\pi^2 T^4}{45} \int d^4 x \sqrt{-\bar g}(\bar g_{00})^{-2}.
\ee
We note that this action is quite different from that which occurs at zero temperature. On the other hand, as argued below, the action associated with the $\ln(T)$ contributions is closely related to the one obtained from Eq. \eqref{eq27}.

It has been shown that the sub-leading $\ln(T)$ contributions of the one-loop Green functions at high temperature, have the same form as the ultraviolet divergent terms at zero temperature \cite{Brandt:1994mc,Brandt:2009ht,Brandt:1998hd}. Consequently, the thermal
$\ln({T^2}/{-k^2})$ terms 
combine with the $\ln(-k^2/\mu^2)$ terms which occur at $T=0$, to yield a $\ln(T^2/\mu^2)$ contribution.
Using an expansion like that shown in Eq. \eqref{eq12},
one can see that such a logarithmic contribution arises from the factor $(\mu/T)^{(4-d)}$.

In general, the Ward identities at finite temperature are different from those at zero temperature due to the fact that the one-point thermal function is non-vanishing.
However, since the Eq. \eqref{eq51} has no logarithmic terms, the Ward identities involving such terms will be the  same as those which occur in the background field method at zero temperature (See, for example, Eqs. \eqref{eq25} and \eqref{eq26}).
From the above properties, it follows that the
high-temperature $\ln(T)$ contributions to the effective action can be directly obtained
by multiplying the effective counter-term Lagrangian \eqref{eq27}
by the thermal factor $(\mu/T)^{(4-d)}$.

We finally note that there is a proposal for an alternative method of quantizing general relativity which leads to a renormalizable and unitary theory \cite{McKeon:2021qhv}.
This involves the introduction of a Lagrangian multiplier (LM) field that restricts the path integrals used to quantize the theory to paths satisfying the classical Euler-Lagrange equations of motion. One finds that such a procedure yields twice the usual one–loop contributions and that all higher order radiative corrections vanish. Thus, by using the (LM) method in quantum gravity, one obtains an exact counter-term Lagrangian which is twice that given in Eq. \eqref{eq1} (see Appendix \ref{appC})
\begin{acknowledgments}
We thank \href{https://www.gov.br/cnpq/}{CNPq} (Brazil) for financial support.
\end{acknowledgments}

\newpage

\appendix

\section{Feynman rules}\label{appA}
\subsection{The action}
We start from the background field action expanded up to second order in the quantum field $h_{\mu\nu}$, which is given by 
\begin{eqnarray}\label{BFaction}
S = \int d^4x \sqrt{-\bar{g}} \left[ {2 \over \kappa^2} \bar{R} 
+{\cal L}_{g}^{(2)}
+{\cal L}_{gf}
+{\cal L}_{gh}
+ {\cal O}(h^3) \right];\;\;\;\; \kappa^2=32\pi G,
\end{eqnarray}
where ${\cal L}^{(2)}_g$, ${\cal L}_{gf}$ and ${\cal L}_{gh}$ are given by Eqs. \eqref{eq4}, \eqref{eq5} and \eqref{eq6}, respectively.
In order to derive the Feynman rules associated with the background field $\bar h_{\mu\nu}$, we use the definition
\begin{subequations}
\be
\bar g_{\mu\nu} = \eta_{\mu\nu} + \kappa \bar h_{\mu\nu}
\ee
as well as the inverse metric up to first order in $\kappa$
\be
\bar g^{\mu\nu} = \eta^{\mu\nu} - \kappa \bar h^{\mu\nu} + {\cal O}(\bar\kappa^2),
\ee
\end{subequations}
which follows from $\bar g^{\mu\alpha} \, \bar g_{\alpha\nu} = \delta^\mu_\nu$. In the present work, we will not need the interactions with more than one background field ${\bar h}_{\mu\nu}$.

For completness, let us also list the usual definitions
\begin{subequations}\label{curv}
\be\label{curv1}
\bar R = \bar g^{\mu\nu}\,\bar R_{\mu\nu},
\ee
\be\label{curv2}
\bar R_{\mu\kappa} = \bar R^\nu_{~\;\mu\nu\kappa},
\ee
\be\label{curv3}
\bar R^\lambda_{~\;\mu\nu\kappa} =
\partial_\nu\,\bar\Gamma^\lambda_{\mu\kappa} -  
\partial_\kappa\,\bar\Gamma^\lambda_{\mu\nu} +  
\bar\Gamma^\sigma_{\mu\kappa}\,\bar\Gamma^\lambda_{\sigma\nu} -  
\bar\Gamma^\sigma_{\mu\nu}\,\bar\Gamma^\lambda_{\sigma\kappa},
\ee
\be\label{curv4}
\bar\Gamma^\lambda_{\mu\nu} = \frac 1 2 \, \bar g^{\lambda\rho}\,
\left(\partial_\nu \bar g_{\rho\mu}+
      \partial_\mu \bar g_{\rho\nu}-
      \partial_\rho \bar g_{\mu\nu}\right)
      =
      \frac \kappa 2 \, \bar \eta^{\lambda\rho}\,
\left(\partial_\nu \bar h_{\rho\mu}+
      \partial_\mu \bar h_{\rho\nu}-
      \partial_\rho \bar h_{\mu\nu}\right) + {\cal O}(\kappa^2),
      \ee
as well as
\end{subequations}

\begin{subequations}\label{covder}
\be\label{covder0}
D_\mu\,\phi = \partial_\mu\,\phi,
\ee
\be\label{covder1}
D_\nu\,\phi_\mu = \partial_\nu\,\phi_\mu - \bar\Gamma^{\alpha}_{\mu\nu}\,\phi_\alpha,
\ee
\be\label{covder2}
D_\sigma\,\phi_{\mu\nu} 
= \partial_\sigma\,\phi_{\mu\nu} 
- \bar\Gamma^{\alpha}_{\mu\sigma}\,\phi_{\alpha\nu}
- \bar\Gamma^{\alpha}_{\nu\sigma}\,\phi_{\mu\alpha},
\ee
\be\label{covder3}
D_\sigma\,\phi_{\mu\nu\cdots} 
= \partial_\sigma\,\phi_{\mu\nu\cdots} 
- \mbox{one}\; \bar\Gamma\; \mbox{for each index} .
\ee
\end{subequations}


Let us consider the leading order expressions for the curvature terms in the second line of Eq. \eqref{eq4}.
Under the change $\bar g_{\mu\nu} \rightarrow \eta_{\mu\nu} + \kappa \bar h_{\mu\nu}$ one finds that, in momentum space,
\be\label{vR}
\sqrt{-\tilde{\bar g}}\tilde{\bar R} 
= \kappa\, q^2 L^{\alpha \beta } \, \tilde{\bar h}_{\alpha\beta} + {\cal O}(\kappa^2),
\ee
where
\be\label{transvL}
L^{\mu\nu}(q) = \frac{q^\mu q^\nu}{q^2} -  \eta^{\mu\nu}.
\ee
Additionally, in the second line of Eq. \eqref{eq4} we have
\be\label{vRmn}
\sqrt{-g} \,\tilde{\bar R}^{\mu\nu}
= 
\frac{\kappa}{2} \left[
L^{\rho\sigma}(q)\,{{q}^{\mu }{q}^{\nu }}
- \frac{q^2}{2} (L^{\mu \sigma}(q) L^{\rho\nu}(q) + L^{\mu \rho}(q) L^{\sigma\nu}(q) )
%
\right] \tilde{\bar h}_{\rho\sigma}(q) + {\cal O}(\kappa^2).
\ee
We point out that the expressions written in terms of combinations of the transverse tensor \eqref{transvL} explicitly exhibit the invariance under the gauge transformation \eqref{GTransf}.

It will also be useful to have the expressions for the following second order variations
\begin{subequations}\label{Rinvariants}
\be\label{RinvA}
\sqrt{-\tilde{\bar g}}\tilde{\bar R}^2
= \kappa^2 q^4 L^{\alpha \beta }(q) L^{\mu \nu }(q)
\tilde{\bar h}_{\mu\nu} \tilde{\bar h}_{\alpha\beta}
+ {\cal O}(\kappa^3),
\ee
\be\label{RinvB}
\sqrt{-\tilde{\bar g}}\tilde{\bar R}_{\mu\nu}\tilde{\bar R}^{\mu\nu}
= \kappa^2 q^4\biggl[
\frac{1}{4} L^{\alpha \beta }(q) L^{\mu \nu}(q)
+\frac{1}{8} L^{\nu \alpha }(q) L^{\mu \beta}(q)
+\frac{1}{8} L^{\mu \alpha }(q) L^{\nu \beta }(q)
\biggr]
\tilde{\bar h}_{\mu\nu} \tilde{\bar h}_{\alpha\beta} + {\cal O}(\kappa^3),
\ee
\be\label{RinvC}
\sqrt{-\tilde{\bar g}}\tilde{\bar R}_{\mu\nu\alpha\beta}\tilde{\bar R}^{\mu\nu\alpha\beta}
= \kappa^2\, q^4
\biggl[
\frac{1}{2} L^{\nu \alpha }(q) L^{\mu \beta}(q)
+\frac{1}{2} L^{\mu \alpha }(q) L^{\nu \beta}(q)
\biggr]
\tilde{\bar h}_{\mu\nu} \tilde{\bar h}_{\alpha\beta} + {\cal O}(\kappa^3)
\ee
\end{subequations}
Notice that the right-hand sides of Eqs. \eqref{Rinvariants} can be viewed as the variations of the corresponding left-hand sides when $\bar g_{\mu\nu}\rightarrow \eta_{\mu\nu}+\kappa\bar h_{\mu\nu}$.
As expected, these variations are compatible with the identity 
\be\label{GB0}
\delta\left(
\sqrt{-\tilde{\bar g}}( \tilde{\bar R}^2 - 4 \tilde{\bar R}_{\mu\nu}\tilde{\bar R}^{\mu\nu}  + \tilde{\bar R}_{\mu\nu\alpha\beta}\tilde{\bar R}^{\mu\nu\alpha\beta})\right) = 0,
\ee
which, in the $x$-space, translates into the vanishing integral of a total derivative (see the Appendix B of \cite{F4}).

\subsection{Propagators for the quantum graviton field $h_{\mu\nu}$ and the ghost field $c$}
The propagator for the field $h_{\mu\nu}$ can be derived from Eqs. \eqref{eq4} and \eqref{eq5}. Replacing all co-variant derivatives by the corresponding partial derivative, we obtain the following quadratic form
\be
i S^{0} = \int d^4 x h_{\mu\nu} Q^{\mu\nu\,\alpha\beta} h_{\alpha\beta},
\ee
where 
\begin{eqnarray}\label{quadratic}
Q^{\mu\nu\,\alpha\beta} &=&
i \left[\frac{1}{\xi}\left({\frac{1}{2} \partial^{\mu } \partial^{\nu } \eta^{\alpha
   \beta }-\partial^{\beta } \partial^{\nu } \eta^{\alpha \mu
   }+\frac{1}{2} \partial^{\alpha } \partial^{\beta } \eta^{\mu \nu
   }-\frac{1}{4} \partial^2 \eta^{\alpha \beta } \eta^{\mu \nu }}\right)
+\partial^{\beta } \partial^{\nu } \eta^{\alpha \mu
   }-\partial^{\alpha } \partial^{\beta } \eta^{\mu \nu }-\frac{1}{2}
   \partial^2 \eta^{\alpha \mu } \eta^{\beta \nu }+\frac{1}{2} \partial^2 \eta^{\alpha \beta }
   \eta^{\mu \nu }\right]
\nonumber \\
&+& \mbox{symmetrizations } \mu\leftrightarrow\nu \mbox{ and } \alpha\leftrightarrow\beta
\nonumber \\
&+& \mbox{permutation } (\mu,\nu) \leftrightarrow (\alpha,\beta) 
\end{eqnarray}
(we have used integration by parts, as usual).

Changing to momentum space ($\partial\rightarrow i p$) and computing $(-\tilde Q^{\mu\nu\alpha\beta}(p))^{-1}$ with the help of the tensor basis in table \ref{tab1} and computer algebra (throughout this work we have used FeynCalc \cite{Shtabovenko:2020gxv}), we obtain
\begin{eqnarray}\label{propquantumh}
{\cal D}^{grav}_{\mu\nu\,\alpha\beta}(p) &=&\frac{i}{p^2+i\epsilon}\Biggl[
 \frac{1}{2}  \left(\eta_{\alpha \nu } \eta_{\beta \mu }
+  \eta_{\alpha \mu } \eta_{\beta \nu }
-\frac{2 \eta_{\alpha \beta } \eta_{\mu \nu}}{d-2}\right)
\nonumber \\
&+&\frac{\xi -1}{2 { p^2}} \biggl(
p_{\beta } p_{\nu } \eta_{\alpha
    \mu }
+ p_{\beta } p_{\mu } \eta_{\alpha
    \nu }
+ p_{\alpha } p_{\nu } \eta_{\beta
    \mu }
+ p_{\alpha } p_{\mu } \eta_{\beta
    \nu }\biggr)\Biggl].
\end{eqnarray}
In the same way, the propagator for the background field $\bar h_{\mu\nu}$ can be found using a gauge fixing term with a gauge parameter $\bar \xi$. This yields a propagator for the $\bar h_{\mu\nu}$ field which is identical to \eqref{propquantumh}, with another gauge fixing \hbox{parameter $\bar \xi$}. 

Similarly, from Eq. \eqref{eq6} we obtain the following expression for the ghost propagator
\be
{\cal D}^{ghost}_{\mu\nu}(p) = -i \frac{\eta^{\mu\nu}}{p^2+i\epsilon}.
\ee

\subsection{The Vertex $\bar h h h$:}
Let us first consider the terms like $DhDh$ in the first line of \eqref{eq4}
and also the gauge fixing contributions in Eq. \eqref{eq5}.
For computer algebra convenience, it helps to first write the general form of such terms as
\be
{\cal L}_{DhDh} = \sqrt{-\bar g}\; T^{\alpha\mu\nu\,\beta\gamma\delta}\bar D_{\alpha} h_{\mu\nu}\bar D_{\beta} h_{\gamma\delta},
\ee
where the tensor $T^{\alpha\mu\nu\,\beta\gamma\delta}$ can be extracted from the first line of Eq. \eqref{eq4}
and from Eq. \eqref{eq5} and is given by
\begin{eqnarray}
T^{\alpha\mu\nu\,\beta\gamma\delta} &=&
 \frac{1}{2} \bar g^{\alpha\beta}\bar g^{\mu\gamma} \bar g^{\nu\delta}
-\frac{1}{2} \bar g^{\alpha\beta}\bar g^{\mu\nu} \bar g^{\gamma\delta}
+\bar g^{\alpha\gamma}\bar g^{\mu\nu} \bar g^{\beta\delta}
-\bar g^{\alpha\delta}\bar g^{\mu\gamma} \bar g^{\nu\beta}
\nonumber \\ &+&
\frac{1}{\xi}\left(
 \bar g^{\alpha\nu}\bar g^{\mu\gamma} \bar g^{\beta\delta}
-\frac{1}{2} \bar g^{\alpha\nu}\bar g^{\mu\beta} \bar g^{\gamma\delta}
-\frac{1}{2} \bar g^{\alpha\gamma}\bar g^{\mu\nu} \bar g^{\beta\delta}
+\frac{1}{4}\bar g^{\alpha\beta}\bar g^{\mu\nu} \bar g^{\gamma\delta}
\right)
\nonumber \\ &=& {H^{(0)}}^{\alpha\mu\nu\beta\gamma\delta} 
+
 \frac{1}{2} {H^{(1)}}^{\alpha\mu\nu\beta\gamma\delta}
-\frac{1}{2} {H^{(1)}}^{\alpha\mu\gamma\beta\nu\delta}
+{H^{(1)}}^{\alpha\mu\beta\gamma\nu\delta}
-{H^{(1)}}^{\alpha\mu\nu\delta\gamma\beta}
\nonumber \\ &+&
\frac{1}{\xi}\left(
 {H^{(1)}}^{\alpha\mu\beta\nu\gamma\delta}
-\frac{1}{2} {H^{(1)}}^{\alpha\mu\gamma\nu\beta\delta}
-\frac{1}{2} {H^{(1)}}^{\alpha\mu\beta\gamma\nu\delta}
+\frac{1}{4} {H^{(1)}}^{\alpha\mu\gamma\beta\nu\delta}
\right)
 + {\cal O}(\kappa^2),
\end{eqnarray}
where
\begin{eqnarray}\label{H0}
{H^{(0)}}^{\alpha\mu\nu\beta\gamma\delta}&\equiv&
 \frac{1}{2} \eta^{\alpha\beta}\eta^{\mu\gamma} \eta^{\nu\delta}
-\frac{1}{2} \eta^{\alpha\beta}\eta^{\mu\nu} \eta^{\gamma\delta}
+\eta^{\alpha\gamma}\eta^{\mu\nu} \eta^{\beta\delta}
-\eta^{\alpha\delta}\eta^{\mu\gamma} \eta^{\nu\beta}
\nonumber \\ &+&
\frac{1}{\xi}\left(
 \eta^{\alpha\nu}\eta^{\mu\gamma} \eta^{\beta\delta}
-\frac{1}{2} \eta^{\alpha\nu}\eta^{\mu\beta} \eta^{\gamma\delta}
-\frac{1}{2} \eta^{\alpha\gamma}\eta^{\mu\nu} \eta^{\beta\delta}
+\frac{1}{4}\eta^{\alpha\beta}\eta^{\mu\nu} \eta^{\gamma\delta}
\right)
\end{eqnarray}
and the tensor ${H^{(1)}}^{\alpha\mu\nu\beta\gamma\delta}$, which arises from the order $\kappa$ contribution in the product of three inverse metric tensors, is given by
\begin{eqnarray}
{H^{(1)}}^{\alpha\mu\nu\beta\gamma\delta}&\equiv &
\kappa\left(-\delta^{\alpha}_{\mu_1}\delta^{\beta}_{\nu_1}  \eta^{\mu\gamma}  \eta^{\nu\delta} 
-\eta^{\alpha\beta}  \delta^{\mu}_{\mu_1}\delta^{\gamma}_{\nu_1}  \eta^{\nu\delta} 
-\eta^{\alpha\beta}  \eta^{\mu\gamma}  \delta^{\nu}_{\mu_1}\delta^{\delta}_{\nu_1}\right)\bar h^{\mu_1\nu_1}.
\nonumber \\ &\equiv &
\kappa\; G^{\mu_1\nu_1\,\alpha\mu\nu\beta\gamma\delta}
\bar h_{\mu_1\nu_1}
\end{eqnarray}
Combining these expressions, we can write part of 
the contributions from ${\cal L}_{DhDh}$ to the $\bar h h h$ interaction vertex
as follows (the subscript $\partial h \partial h$ indicates that this part originates from terms which have two derivatives acting on the quantum field $h_{\mu\nu}$)
\begin{eqnarray}
{\cal V}_{\partial h\partial h}^{\mu_1\nu_1\,\mu_2\nu_2\,\mu_3\nu_3}(p_1,p_2,p_3)&=&
  -i\kappa
  \left[\frac{1}{2} \eta^{\mu_1\nu_1} {H^{(0)}}^{\alpha\mu\nu\beta\gamma\delta}
+    \frac{1}{2} G^{\mu_1\nu_1\,\alpha\mu\nu\beta\gamma\delta}
-\frac{1}{2} G^{\mu_1\nu_1\,\alpha\mu\gamma\beta\nu\delta}
+G^{\mu_1\nu_1\,\alpha\mu\beta\gamma\nu\delta}
-G^{\mu_1\nu_1\,\alpha\mu\nu\delta\gamma\beta}
\right.
\nonumber \\ &+&
\frac{1}{\xi}\left.\left(
 G^{\mu_1\nu_1\,\alpha\mu\beta\nu\gamma\delta}
-\frac{1}{2} G^{\mu_1\nu_1\,\alpha\mu\gamma\nu\beta\delta}
-\frac{1}{2} G^{\mu_1\nu_1\,\alpha\mu\beta\gamma\nu\delta}
+\frac{1}{4} G^{\mu_1\nu_1\,\alpha\mu\gamma\beta\nu\delta}
\right)
\right]
\nonumber \\
&&  \delta^{\mu_2}_{\mu} \delta^{\nu_2}_{\nu} \delta^{\mu_3}_{\gamma} \delta^{\nu_3}_{\delta}
 \; {p_2}_\alpha {p_3}_\beta
+ \mbox{ symmetrizations } \mu_i\leftrightarrow\nu_i
\nonumber \\
&+& \mbox{ permutation } (\mu_2,\nu_2,p_2) \leftrightarrow (\mu_3,\nu_3,p_3),
\end{eqnarray}
where the first term arises from the order $\kappa$ contribution of $\sqrt{-\bar g}$. 
In this form, one can easily input the definition of $G^{\mu_1\nu_1\alpha\mu\nu\beta\gamma\delta}$ in an computer algebra system and obtain the full expression for this part of the $\bar h h h$ interaction vertex.

There is also another contribution from the first order terms of the co-variant derivatives. Using the expression \eqref{covder2}, the order $\kappa$ contribution can be written as
\begin{eqnarray}
{\cal V}_{\partial \bar h\partial h}^{\mu_1\nu_1\,\mu_2\nu_2\,\mu_3\nu_3}(p_1,p_2,p_3)&=&
i \kappa \;\Biggr\{\Biggl[\frac{1}{2}
  \Biggl( \delta_{\mu}^{ \mu_3} \delta_{\nu}^{\nu_3} {p_1}_{\gamma }
   {p_3}_{\alpha } \delta_{\beta}^{ \nu_1} \delta_{\delta}^{ \mu_2} \eta^{\mu_1\nu_2}
   + \delta_{\mu}^{\mu_3} \eta^{\nu_1\nu_2} \delta_{\nu}^{\nu_3} {p_1}_{\beta } {p_3}_{\alpha }
   \delta_{\gamma}^{ \mu_1} \delta_{\delta}^{ \mu_2}
- \delta_{\mu}^{\mu_3} \delta_{\nu}^{\nu_3}
   {p_1}^{\nu_2} {p_3}_{\alpha } \delta_{\beta}^{ \mu_1} \delta_{\gamma}^{ \mu_2} \delta_{\delta}^{ \nu_1}
   \Biggr)
   \nonumber \\  \qquad\;\;\; &+&
    \gamma \leftrightarrow \delta\Biggr]
+ (\alpha,\mu,\nu) \leftrightarrow(\beta,\gamma,\delta)
   \Biggr\} {H^{(0)}}^{\alpha\mu\nu\beta\gamma\delta}
+ \mbox{ symmetrizations } \mu_i\leftrightarrow\nu_i
\nonumber \\
&+& \mbox{ permutation } (\mu_2,\nu_2,p_2) \leftrightarrow (\mu_3,\nu_3,p_3),
\end{eqnarray}

The third type of contribution arises from the curvature terms in the second line of Eq. \eqref{eq4}. 
With the help of Eqs. \eqref{vR} and \eqref{vRmn} we obtain the following result
\begin{eqnarray}
{\cal V}_{\partial \bar h \partial \bar h}^{\mu_1\nu_1\,\mu_2\nu_2\,\mu_3\nu_3}(p_1,p_2,p_3)&=&
\kappa \Big\{p_1^2 \, L^{\mu_1\nu_1}(p_1)
\left(
 \frac{1}{4} \eta^{\mu_2\nu_2} \eta^{\mu_3\nu_3} - \frac{1}{2}\eta^{\mu_2\mu_3}\eta^{\nu_2\nu_3}\right)
 \nonumber \\ &+& 
 \frac{1}{2}\left[
L^{\mu_1\nu_1}(p_1) {p_1}^\mu {p_1}^\nu - \frac{p_1^2}{2}\left(
L^{\mu\mu_1}(p_1) L^{\nu\nu_1}(p_1) + L^{\nu\mu_1}(p_1) L^{\mu\nu_1}(p_1)
\right)
\right]
\nonumber\\ && 
 \left(
2 \eta^{\nu_3\mu_2} \delta_{\mu}^{\nu_2} \delta_{\nu}^{\mu_3}-\eta^{\mu_2\nu_2}\delta_\mu^{\mu_3}\delta_\nu^{\nu_3}
\right)
\Big\}
+ \mbox{ symmetrizations } \mu_i\leftrightarrow\nu_i
\nonumber \\
&+& \mbox{ permutation } (\mu_2,\nu_2,p_2) \leftrightarrow (\mu_3,\nu_3,p_3).
\end{eqnarray}
In the previous expressions we have not performed some index contractions, because this can be easily done using FeynCalc \cite{Shtabovenko:2020gxv}. The total result for the $\bar h_{\mu_1\nu_1} h_{\mu_2\nu_2} h_{\mu_3\nu_3}$ interactions vertex is
\be\label{VgravTot}
{\cal V}_{grav}^{\mu_1\nu_1\,\mu_2\nu_2\,\mu_3\nu_3}(p_1,p_2,p_3)=
   {\cal V}_{\partial h \partial h}^{\mu_1\nu_1\,\mu_2\nu_2\,\mu_3\nu_3}(p_1,p_2,p_3)+
   {\cal V}_{\partial \bar h \partial h}^{\mu_1\nu_1\,\mu_2\nu_2\,\mu_3\nu_3}(p_1,p_2,p_3)+
   {\cal V}_{\partial \bar h \partial \bar h}^{\mu_1\nu_1\,\mu_2\nu_2\,\mu_3\nu_3}(p_1,p_2,p_3).
\ee

\subsection{The vertex $c^\star \bar h c$}
It is convenient to write the ghost Lagrangian given in Eq. \eqref{eq6} as follows
\begin{eqnarray}\label{LghostN}
 {\cal L}_{gh} &=& \sqrt{-\bar g}c^\star_{\mu}
\left(\bar g^{\beta\lambda} \bar g^{\mu\nu} \bar D_\beta \bar D_\lambda - \bar R^{\mu\nu}\right)c_\nu
\nonumber \\
&=& c^\star_\mu \partial^2 c^\mu
+ \kappa \Biggl\{\frac{1}{2} \bar h^\lambda_\lambda c^\star_{\mu} \partial^2 c^\mu
-c^\star_{\mu} \left(\bar h^{\beta\lambda}\eta^{\mu\nu} + \bar h^{\mu\nu}\eta^{\beta\lambda}\right) \partial_\beta\partial_\lambda c_\nu
\nonumber \\
&+&\frac{1}{2} \eta^{\mu\nu}\eta^{\beta\lambda}
\biggl[
(\partial_\beta c^\star_\mu) \eta^{\rho\sigma}
\left(
\partial_\lambda\bar h_{\sigma\nu}+\partial_\nu\bar h_{\sigma\lambda}-\partial_\sigma\bar h_{\nu\lambda}\right)c_{\rho}
\nonumber \\
&&\;\;\;\;\;\;\;\;\;\;\;\;\;-c^\star_\mu\eta^{\rho\sigma}
\left(
\partial_\beta\bar h_{\sigma\lambda}+\partial_\lambda\bar h_{\sigma\beta}-\partial_\sigma\bar h_{\lambda\beta}\right)\partial_\rho c_{\nu}
\nonumber \\
&&\;\;\;\;\;\;\;\;\;\;\;\;\;-c^\star_\mu\eta^{\rho\sigma}
\left(
\partial_\beta\bar h_{\sigma\nu}+\partial_\nu\bar h_{\sigma\beta}-\partial_\sigma\bar h_{\nu\beta}\right)\partial_\lambda c_{\rho}
\biggr]\Biggr\} - c^\star_\mu \bar R^{\mu\nu} c_\nu .
\end{eqnarray}
From \eqref{LghostN} we obtain the following momentum space interaction vertex $\bar h_{\mu_1\nu_1} c^\star_\mu c_\nu$
\begin{eqnarray}
{\cal V}_{ghost}^{\mu_1\nu_1\,\mu\nu}(p_1,p_2,p_3)&=& i\kappa\bigg\{
-\frac{1}{2} p_3^2 \eta^{\mu\nu} \eta^{\mu\nu} 
+p_3^{\mu_1}p_3^{\nu_1}\eta^{\mu\nu} + p_3^2 \eta^{\mu\mu_1} \eta^{\nu\nu_1} 
\nonumber \\ &+&\frac{1}{2}\left(
p_1^{\nu} p_2^{\nu_1} \eta^{\mu\mu_1}-p_1^{\mu} p_2^{\nu_1} \eta^{\nu\mu_1}
-p_1\cdot p_3 \eta^{\mu\nu_1} \eta^{\nu\mu_1}
\right)
\nonumber \\ &+& \frac{1}{2}\left(
2 \, p_3^{\mu_1} p_1^{\nu_1} \eta^{\mu\nu}-p_1\cdot p_3 \eta^{\mu\nu}\eta^{\mu_1\nu_1}
\right)
\nonumber \\ &-&\frac{1}{2}\left(
p_1^{\nu} p_2^{\nu_1} \eta^{\mu\mu_1}-p_1^{\mu} p_2^{\mu_1} \eta^{\nu\nu_1}
-p_1\cdot p_3 \eta^{\mu\nu_1} \eta^{\nu\mu_1}
\right)
\nonumber \\
&-&\frac{1}{2}\left[
p_1^{\mu_1}\left(
p_1^{\mu}\eta^{\nu\nu_1}+p_1^{\nu}\eta^{\mu\nu_1}\right) - p_1^{\mu} p_1^{\nu} \eta^{\mu_1\nu_1} - p1^2\eta^{\mu\mu_1}\eta^{\nu\nu_1}
\right]
\bigg\}
\nonumber \\
&+& \mbox{ symmetrizations } \mu_1\leftrightarrow\nu_1 ,
\end{eqnarray}
where in the next to the last line we have used Eq. \eqref{vRmn}.

\subsection{Ward identities}
When we use $g_{\mu\nu} = \eta_{\mu\nu}+\kappa h_{\mu\nu}$ in the Einstein-Hilbert action, we find 
identities which relate ${\cal V}_{\mu _1\nu _1\cdots \mu _{n-1}\nu _{n-1}}^{n-1}$ to
${\cal V}_{\mu _1\nu _1\cdots \mu _{n}\nu _{n}}^{n}$. This is a direct consequence of the invariance of the classical action under the transformation
\be\label{GTransf}
\kappa \delta h_{\mu\nu} =  D_\mu\,\omega_\nu +  D_\nu\,\omega_\mu. 
\ee
A straightforward analysis leads to
$$
-\left( \eta ^{\mu_1\sigma }\delta_\lambda ^{\nu_1}+\eta ^{\nu_1\sigma
}\delta_\lambda ^{\mu_1}\right)\, p_\sigma ^1\;
{\cal V}_{\mu_1\nu_1\,\mu_2\nu_2\,\mu_3\nu_3\cdots \mu_n\nu_n}^n
\left( p^1,p^2,p^3,\cdots ,p^n\right) = 
$$
$$
\kappa\,W^{\mu_1\nu_1}_{\mu_2\nu_2\lambda}(p^1,p^2)\;
{\cal V}_{\mu_1\nu_1\,\mu_3\nu_3\cdots \mu_n\nu_n}^{n-1}\left(
p^1+p^2,p^3,\cdots ,p^n\right) 
$$
$$
+ \kappa\,W^{\mu_1\nu_1}_{\mu_3\nu_3\lambda}(p^1,p^3)\;
{\cal V}_{\mu_1\nu_1\,\mu_2\nu_2\,\mu_4\nu_4\cdots \mu_n\nu_n}^{n-1}\left(
p^1+p^3,p^2,p^4\cdots ,p^n\right) 
$$
$$
+ \cdots + 
$$
\be\label{Gward1}
+ \kappa\,W^{\mu_1\nu_1}_{\mu_n\nu_n\lambda}(p^1,p^n)\;
{\cal V}_{\mu_1\nu_1\,\mu_2\nu_2\cdots \mu_{n-1}\nu_{n-1}}^{n-1}\left(
p^1+p^n,p^2,\cdots ,p^{n-1}\right) , 
\ee
where
\be\label{defW1}
W^{\mu\nu}_{\mu_i\nu_i\lambda}(p^1,p^i) \equiv
\frac 1 2 \left(\delta^\mu_{\mu_i}\,\eta_{\nu_i\lambda}+
                \delta^\mu_{\nu_i}\,\eta_{\mu_i\lambda}
          \right){p^1}^{\nu}+
\frac 1 2 \delta^{\mu}_{\mu_i}\,\delta^{\nu}_{\nu_i}\,p^i_\lambda
+\mu \leftrightarrow \nu
\ee
and   $p^1+p^2+\cdots +p^n=0$.
Using the results for the vertex in Eq. \eqref{VgravTot} and the quadratic term in Eq. \eqref{quadratic} we have verified \eqref{Gward1} for $n=2,\; 3$.

\section{The background field self-energy}\label{appB}
The one-loop contributions to the two-point function $\langle \bar h\bar h \rangle$ are given
by the Feynman diagrams of Fig. (\ref{fig1}).
Since we are using dimensional regularization \cite{tHooft:1972tczBollini:1972ui},
the one-loop contributions with a single quartic coupling (see Fig. (\ref{fig1}-c)) vanish. 
All we need are the Feynman rules for the cubic vertices derived in the appendix \ref{appA}.
\begin{table}[b!]
\begin{center}
\begin{tabular}{c l}\hline \hline \\
${\cal T}^{1} _{\mu \nu \,\alpha \beta} (u,k)=$&$  k_\mu k_\nu k_\alpha  k_\beta$\\ & \\
${\cal T}^2 _{\mu \nu \,\alpha \beta} (u,k)=$&$ \eta_{\mu \nu} \eta_{\alpha\beta}$ \\ & \\
${\cal T}^3 _{\mu \nu \, \alpha \beta}(u,k)=$&$  \eta _{\mu \alpha} \eta_{\nu \beta} + \eta _{\mu \beta} \eta_{\nu \alpha}$  \\ & \\
${\cal T}^4 _{\mu \nu \,\alpha \beta} (u,k)=$&$ \eta_{\mu \nu} k_\alpha k_\beta + \eta_{\alpha \beta} k_\mu k_\nu$ \\ & \\
${\cal T}^5 _{\mu \nu \, \alpha \beta}(u,k)=$&$ \eta_{\mu \alpha} k_\nu k_\beta +
\eta_{\mu \beta} k_\nu k_\alpha + \eta_{\nu \alpha} k_\mu k_\beta +
\eta_{\nu \beta} k_\mu k_\alpha$ 
\\  \\ \hline \hline 
\end{tabular}\caption{The five independent tensors built from 
$\eta_{\mu\nu}$ and $k_\mu$, satisfying the symmetry conditions 
${\cal T}^{i}_{\mu\nu\,\alpha\beta} (k)= {\cal T}^{i}_{\nu\mu\,\alpha\beta}(k)= 
{\cal T}^{i}_{\mu\nu\,\beta\alpha} (k)= {\cal T}^{i}_{\alpha\beta\,\mu\nu}(k)$.}\label{tab1}
\end{center}
\end{table}

After loop integration,  the result can only depend (by covariance) on the five tensors shown 
in table \ref{tab1}, so that each diagram in figure (\ref{fig1}) can be written as 
\begin{equation}\label{eq2a}
\Pi^I_{\mu\nu\,\alpha\beta}(k) =  \sum_{i=1}^{5}  {\cal T} ^i_{\mu\nu\, \alpha\beta}(k) 
C^I_i(k) ; \;\;\; I=\mbox{ghost or graviton loop} .
\end{equation}
The coefficients $C^I_i$ can be obtained solving the following system of five algebraic equations
\be
\sum_{i=1}^5 {\cal T}^i_{\mu\nu\,\alpha\beta}(k) {\cal T}^j{}^{\mu\nu\,\alpha\beta}(k) C^I_i(k) =
\Pi^I_{\mu\nu\alpha\beta}(k) {\cal T}^j{}^{\mu\nu\,\alpha\beta}(k) \equiv J^I{}^j(k); \;\;
j=1,\dots,5 .
\ee
Using the Feynman rules for $\Pi^I_{\mu\nu\,\alpha\beta}(k)$ (the vertices and propagators are given in the Appendix \ref{appA})
the integrals on the right hand side have the following form
\be
J^I{}^j(k) = \int \frac{d^d p}{(2 \pi)^d}  s^I{}^j(p,q,k).
\ee
where $q=p+k$; $p$ is the loop momentum, $k$ is the external momentum and $s^I{}^j(p,q,k)$ are
scalar functions. Using the relations 
\begin{subequations}
\begin{eqnarray}
p\cdot k = (q^2 - p^2 - k^2)/2, \\
q\cdot k = (q^2 + k^2 - p^2)/2, \\
p\cdot q = (p^2 + q^2 - k^2)/2, 
\end{eqnarray}
\end{subequations}
the scalars  $s^I{}^j(p,q,k)$ can be reduced to combinations of powers of $p^2$ and $q^2$. As a result, the
integrals $J^I{}^j(k)$  can be expressed in terms of combinations of the following well known integrals 
\begin{equation}
I^{ab} \equiv \mu^{4-d}
\int \frac{d^d p}{i(2 \pi)^d} \frac{1}{(p^2)^a (q^2)^b} =  \frac{(k^2)^{d/2-a-b}}{(-\mu^2)^{d/2-2}(4\pi)^{d/2}}
\frac{\Gamma(a+b-d/2)}{\Gamma(a) \Gamma(b)} \frac{\Gamma(d/2-a) \Gamma(d/2-b)}{\Gamma(d-a-b)} 
\end{equation}
(this has also been considered in \cite{Chetyrkin:1980pr}).
The only non-vanishing (i.e. non tadpole) integrals are the ones with both $a> 0$ and $b> 0$. 
%
For a general gauge parameter, $\xi \neq 1$ 
the diagram in Fig. (\ref{fig1}-a) involves the following three kinds of integrals
(the ghost loop only involves $I^{11}$)
\begin{subequations}\label{intregd}
\begin{eqnarray}
I^{11} & = & \frac{(-k^2/\mu^2)^{d/2-2}}{2^d\pi^{d/2}}
\frac{\Gamma \left(2-\frac{d}{2}\right) \Gamma \left(\frac{d}{2}-1\right)^2}{\Gamma (d-2)} \\
I^{12} & = & I^{21} = \frac{(3-d) I^{11}} {k^2} \\ 
I^{22} & = & \frac{(3-d) (6-d) I^{11} } {k^4} ,
\end{eqnarray}
\end{subequations}
where we have employed the basic property of the $\Gamma$-function $x \Gamma(x) = \Gamma(x+1)$. For $d=4-2\epsilon$ we obtain
\be\label{I11div}
\left.I^{11}\right|_{4-2\epsilon} \equiv I^{div} = \frac{1}{16\pi^2}\left[
\frac{1}{\epsilon} - \ln\left(-\frac{k^2}{4\pi\mu^2}\right) -\gamma + 2\right] + {\cal O}(\epsilon),
\ee
where $\gamma\approx 0.5772$ is the Euler–Mascheroni constant.

A straightforward computer algebra code can now be setup in order implement the steps above described
and to obtain the structures
$C^{{ghost}}_i$ and $C^{{graviton}}_i$. The results are the following
\begin{subequations}\label{Cghost}
\begin{eqnarray}
C_1^{ghost} &=& (C^{ghost}_2+ 2 C^{ghost}_3)\frac{1}{k^4}  \\
C_2^{ghost} &=& -\frac{(d (d (d+10)-10)-20) }{16 \left(d^2-1\right)}  \kappa^2 \, k^4 I^{11}\\
C_3^{ghost} &=& \frac{((3-2 d) d+6) }{16 \left(d^2-1\right)} \kappa^2\, k^4 I^{11}\\
C_4^{ghost} &=& - C_2^{ghost} \frac{1}{k^2} \\
C_5^{ghost} &=& - C_3^{ghost} \frac{1}{k^2}.
\end{eqnarray}
\end{subequations}

In the background field approach, we expect that both the ghost loop and the graviton loop satisfy 
the Ward identity (see Eq. \eqref{Gward1})
\be\label{Pitransv}
(\delta_{\mu}^{\rho} k_\nu + \delta_{\nu}^{\rho} k_\mu)  \Pi^{\mu\nu\,\alpha\beta} = 0
\ee
which follows from the diffeomorphism invariance under \eqref{GTransf}. 
This identity implies that the $C^I_i$ in Eq. \eqref{eq2a}  are not independent.
Substituting Eq. \eqref{eq2a} into \eqref{Pitransv}, a simple algebraic manipulation yields the relations 
\be\label{WardCs}
C^I_1  =  \frac{C^I_2 + 2 C^I_3}{k^4};\;\;\; 
C^I_4  = - \frac{C^I_2}{k^2};\;\;\; C^I_5 = -\frac{C_3^I}{k^2}, 
\ee
which is exactly the relations found in \eqref{Cghost}. Therefore, the ghost loop contribution is transverse. 

The result for the graviton loop is also transverse with the the two independent structure constants given by
\begin{subequations}\label{Cgraviton}
\begin{eqnarray}
C^{{graviton}}_2 &=&
\left[\frac{d (d (d (9 d-52)+74)+68)-96}{64 (d-2)(d-1)}+
\frac{d (d ((d-8) d+20)-14)(\xi -1)}{8 (d-2) (d-1)}
\right. \nonumber \\
&+&\left. \frac{(d-2) d ((d-2) d-2) (\xi -1)^2}{32 (d-1)}
\right] \kappa^2 \,k^4 I^{11} \\
C^{{graviton}}_3 &=&
\left[
\frac{d (d (4 d+5)-98)+112}{64 (d-2) (d-1)}
+\frac{d ((d-5) d+5) (\xi -1)}{8 (d-2)(d-1)} 
+\frac{(d-2) d (\xi -1)^2}{32 (d-1)}
\right] \kappa^2 \, k^4 I^{11} 
\end{eqnarray}
\end{subequations}
(the other three constants can be obtained from \eqref{WardCs}).

A more compact way to write a general tensor which obeys the gauge invariance constraints in Eq. \eqref{WardCs}, is in terms of the tensor $L^{\mu\nu}$ defined in Eq. \eqref{transvL}.
Using this tensor basis, we can write the self-energy as follows
\begin{eqnarray}\label{PiLL}
\Pi^{\mu_1\nu_1\,\mu_2\nu_2}(k) &=& 
(C^{ghost}_2 + C_2^{graviton}) L^{\mu_1\nu_1}(k) L^{\mu_2\nu_2}(k)
\nonumber \\
&+& (C^{ghost}_3 + C_3^{graviton})\left[ L^{\mu_1\mu_2}(k) L^{\nu_1\nu_2}(k)  +  L^{\mu_1\nu_2}(k) L^{\mu_2\nu_1}(k) \right]
\end{eqnarray}

Let us now relate these results with the counter-term 
\begin{eqnarray}\label{Ldiv}
   {\cal L}_{div} =  C_{div}\left(C_{\bar R^2}  \sqrt{-\bar g}
  \bar R^2 + C_{\bar R_{\mu\nu}\bar R^{\mu\nu}} \sqrt{-\bar g} \bar R_{\mu\nu} \bar R^{\mu\nu}\right)
\end{eqnarray}
(notice that we have consistently obtained the Feynman rules from $i S$).
From Eqs. \eqref{RinvA} and \eqref{RinvB}, the momentum space expression corresponding to \eqref{Ldiv} can be written in terms of $L^{\mu\nu}(k)$ as follows 
\begin{eqnarray}\label{LdivN}
   \tilde{\cal L}_{div} &=&    \kappa^2 k^4 C_{div}\Biggr[C_{\bar R^2}
    L^{\alpha \beta }(k) L^{\mu \nu }(k)
\nonumber \\ 
&+&C_{\bar R_{\mu\nu} \bar R^{\mu\nu}}\left(\frac{1}{4} L^{\alpha \beta }(k) L^{\mu \nu}(k)
+\frac{1}{8} L^{\nu \alpha }(k) L^{\mu \beta}(k)
+\frac{1}{8} L^{\mu \alpha }(k) L^{\nu \beta }(k)\right)
    \Biggr]   \tilde{\bar h}_{\mu\nu}(k) \tilde{\bar h}_{\alpha\beta}(k).
\end{eqnarray}
Taking the functional derivative of $\tilde {\cal L}_{div}$, we obtain
\begin{eqnarray}\label{LdivNDD}
\frac 1 2  \frac{\delta^2 \tilde{\cal L}_{div}}
    {\delta \tilde{\bar h}_{\mu_1\nu_1}(k) \delta \tilde{\bar h}_{\mu_2\nu_2}(k)}
  &=&     \kappa^2 k^4 C_{div}\Biggr[C_{\bar R^2}
    L^{\mu_1 \nu_1 }(k) L^{\mu_2 \nu_2 }(k)
\nonumber \\ 
&+&C_{\bar R_{\mu\nu} \bar R^{\mu\nu}}\left(\frac{1}{4} L^{\mu_1 \nu_1 }(k) L^{\mu_2 \nu_2}(k)
+\frac{1}{8} L^{\nu_2 \mu_1 }(k) L^{\mu_2 \nu_1}(k)
+\frac{1}{8} L^{\mu_2 \mu_1 }(k) L^{\nu_1 \nu_1 }(k)\right)
    \Biggr]   
\end{eqnarray}
Comparing Eq. \eqref{PiLL} with \eqref{LdivNDD} and using the expressions given in Eqs. \eqref{Cghost} and \eqref{Cgraviton}, we obtain
\begin{subequations}\label{B17}
\begin{eqnarray}
C_{div} C_{\bar R^2} &=& I^{11}
\Biggl[\frac{d (d (d (d (9 d-55)-12)+392)-56)-384}{128 (d-2) (d^2-1)}
+\frac{(d-4) d ((d-6) d+6)}{16 (d-2) (d-1)}(\xi -1)
\nonumber \\
&+&\frac{d \left((d-4) d^2+8\right)}{64 (d-1)} (\xi -1)^2 \Biggr] 
\nonumber \\
&=& I^{div} 
\left(\frac{1}{120} + \frac{1}{6} (\xi -1)^2 \right) + \cdots 
\end{eqnarray}
and
\begin{eqnarray}\label{B20}
  C_{div} C_{\bar R_{\mu\nu} \bar R^{\mu\nu}} & = &  I^{11}
\left[  \frac{4 d^4+d^3-65 d^2+14 d+64}{16 (d-2) (d^2-1)} +
  \frac{d \left(d^2-5 d+5\right)}{2 (d-2) (d-1)} (\xi-1) +
  \frac{(d-2) d}{8 (d-1)} (\xi-1)^2 \right] 
  \nonumber \\ 
  &=& I^{div}  
  \left(
\frac{7}{20}+\frac{\xi (\xi -1)}{3} \right) + \cdots, 
\end{eqnarray}
\end{subequations}
where $I^{div}$ is given by Eq. \eqref{I11div}
and the $\cdots$ represent all the finite terms in the limit $\epsilon\rightarrow 0$.
Eqs. \eqref{B17} agree with the well known result in the gauge $\xi=1$ \cite{F4}.

Finally, using Eqs. \eqref{PiLL} and \eqref{propquantumh} we obtain the following one-loop correction to the propagator of the background field
\begin{eqnarray}\label{1loopProp}
\bar {\cal D}^{\alpha\beta\,\rho\sigma} \Pi_{\rho\sigma\,\lambda\delta} \bar{\cal D}^{\lambda\delta\,\mu\nu} &=& 
-I^{11} \Biggl\{
  \Biggl[
\frac{4 d^4+d^3-65 d^2+14 d+64}{64 (d-2) (d^2-1)} +
\frac{d \left(d^2-5 d+5\right)}{8 (d-2) (d-1)} (\xi-1) +
\frac{(d-2) d}{32 (d-1)}(\xi-1)^2
\Biggr]
\nonumber \\ &&
\qquad\qquad
\left(\eta^{\alpha \nu } \eta^{\beta \mu}+\eta^{\alpha \mu } \eta^{\beta \nu }\right)
\nonumber \\ &+&
\Biggl[
\frac{d^4-23 d^3+80 d^2+4 d-64}{64 (d-2)^2 \left(d^2-1\right)}
-\frac{(d-4) d}{8 (d-2) (d-1)} (\xi-1)
+\frac{(d-2) d}{32 (d-1)} (\xi-1)^2
\Biggr] \eta^{\alpha\beta}  \eta^{\mu\nu} 
\nonumber \\
   &+& \mbox{ terms containing } k^\alpha,\,k^\beta,\,k^\mu,\, k^\nu \Biggr\} ,
\end{eqnarray}
where $\bar{\cal D}^{\mu\nu \,\alpha\beta}$ is obtained from Eq. \eqref{propquantumh} replacing $\xi$ by $\bar\xi$. The above expression reduces, for $d = 4 - 2 \epsilon$, to the result 
\begin{eqnarray}\label{B18a}
\left.\bar {\cal D}^{\alpha\beta\,\rho\sigma} \Pi_{\rho\sigma\,\lambda\delta} \bar{\cal D}^{\lambda\delta\,\mu\nu}\right|_{d=4-2\epsilon} &=& 
- \frac{I^{div}}{2}   
\biggl\{
\left[\frac{21}{120}+\frac{\xi(\xi-1)}{6}\right]
\left(\eta^{\alpha \nu } \eta^{\beta \mu}+\eta^{\alpha \mu } \eta^{\beta \nu }\right)
+\left[\frac{1}{120}+\frac{(\xi-1)^2}{6}\right]\eta^{\alpha \beta } \eta^{\mu \nu }
\nonumber \\
&+& \mbox{ terms containing } k^\alpha,\,k^\beta,\,k^\mu,\, k^\nu \Biggr\} + \cdots,
\end{eqnarray}
which agrees with the known result in the gauge $\xi=1$
\cite{Donoghue:1993eb,F15}.
%
In Sec. \ref{sec3}
we employ this result to compute the quantum correction to the Newtonian potential.

\section{Using a Lagrange Multiplier Field to eliminate radiative corrections beyond one-loop}\label{appC}
It has been established that by supplementing the classical Lagrangian with a term in which a Lagrange multiplier (LM) field is used to ensure that the classical equations of motion are satisfied, the radiative corrections to the classical action are restricted to one-loop order. These one-loop corrections are twice those arising from the classical action alone. (See Ref. \cite{McKeon:2021qhv} and references therein.) This has made it possible to absorb all divergences arising from the Einstein-Hilbert action (both by itself and when interacting with a scalar field) into the LM field, leaving a finite result that is consistent with unitarity.

We now will summarize the consequence of including such a LM field when we do not specify the gauge fixing parameter $\xi$. Starting from the Einstein-Hilbert Lagrangian of Eq. \eqref{eq2}, we find that since
\be\label{c1}
\delta\int d^4 x {\cal L}_g = -\frac{2}{\kappa^2}\int d^4 x\sqrt{-g}\;
\delta g_{\mu\nu}\left(R^{\mu\nu}-\frac{1}{2} g^{\mu\nu} R\right),
\ee
we introduce a LM field $\lambda^{\mu\nu}$ so that the Einstein-Hilbert Lagrangian is supplemented by
\be\label{c2}
{\cal L}_\lambda = \left(-\frac{2}{\kappa^2}\right)\sqrt{-g} \lambda^{\mu\nu}
\left(R_{\mu\nu}-\frac{1}{2} g_{\mu\nu}R\right).        
\ee
The LM field $\lambda^{\mu\nu}$ is now split into a background part $\bar\lambda^{\mu\nu}$ and a quantum fluctuation $\sigma^{\mu\nu}$
\be\label{c3}
\lambda^{\mu\nu} = \bar\lambda^{\mu\nu} + \sigma^{\mu\nu},
\ee
and now $\int d^4 x ({\cal L}_g  +  {\cal L}_\lambda)$ is invariant under two gauge transformations \cite{McKeon:2021qhv},
\begin{subequations}\label{c4}
\be\label{c4a}
\delta h_{\mu\nu} = \frac{1}{\kappa}\left(\bar D_\mu \theta_\nu + \bar D_\nu \theta_\mu\right)
+\theta^\lambda \bar D_\lambda h_{\mu\nu}
+h_{\mu\lambda}\bar D_\nu \theta^\lambda+h_{\nu\lambda}\bar D_\mu \theta^\lambda ,
\ee
\be\label{c4b}
\delta\sigma_{\mu\nu}=
\left(\bar\lambda_{\mu\lambda}+\sigma_{\mu\lambda}\right)\bar D_\nu\theta^\lambda+
\left(\bar\lambda_{\nu\lambda}+\sigma_{\nu\lambda}\right)\bar D_\mu\theta^\lambda+
\theta^\lambda D_\lambda \left(\bar\lambda_{\mu\nu}+\sigma_{\mu\nu}\right) 
\ee
and
\be\label{c4c}
\delta\sigma_{\mu\nu}=
\left(\bar D_\mu \chi_\nu + \bar D_\nu \chi_\mu + \chi^\lambda \bar D_\lambda\sigma_{\mu\nu}\right)+
\sigma_{\mu\lambda} \bar D_\nu \chi^\lambda + \sigma_{\nu\lambda} \bar D_\mu \chi^\lambda ,
\ee 
\end{subequations}
with $\bar g_{\mu\nu}$ and $\bar\lambda^{\mu\nu}$ held constant. Breaking the gauge invariances of Eqs. \eqref{c4} while preserving gauge invariance of the background fields $\bar g_{\mu\nu}$ and $\bar\lambda^{\mu\nu}$ involves replacing Eqs. \eqref{eq5} and \eqref{eq6} by \cite{McKeon:2021qhv}
\begin{subequations}\label{c5}
\begin{eqnarray}\label{c5a}
  {\cal L}_{gf\lambda} &=& \frac{1}{\xi}\sqrt{-\bar g}\Bigg[
    \left( \bar D^{\nu} h_{\mu \nu} - {1 \over 2} \bar D_{\mu} h^\alpha_\alpha \right)
    \left( \bar D_{\sigma}h^{\mu\sigma} - {1 \over 2} \bar D^{\mu} h^\beta_\beta \right)
    \nonumber \\ &&
     \;\;\;\;\;\;\;\;
   +2\left( \bar D^{\nu} \sigma_{\mu \nu} - {1 \over 2} \bar D_{\mu} \sigma^\alpha_\alpha \right)
    \left( \bar D_{\sigma}h^{\mu\sigma} - {1 \over 2} \bar D^{\mu} h^\beta_\beta \right)
    \Bigg]
\end{eqnarray}
\begin{equation}\label{c5b}
 {\cal L}_{gh\lambda} = \sqrt{-\bar g}\left[
  c^{* \mu} \left( \bar D_{\lambda} \bar D^{\lambda} \bar{g}_{\mu \nu} - \bar R_{\mu \nu}\right) d^{\nu}
+ d^{* \mu} \left( \bar D_{\lambda} \bar D^{\lambda} \sigma_{\mu \nu} - \bar R_{\mu \nu}\right) c^{\nu}\right],
\end{equation}
\end{subequations}
where ${d^\star}^\mu$ and $d^\nu$ are additional Grassmann ghost fields.

In the path integral for the one-particle irreducible Feynman diagrams, one can do explicitly the path integral over the fields $h_{\mu\nu}$ and $\sigma^{\mu\nu}$ leaving one with \cite{McKeon:2021qhv}
\begin{eqnarray}\label{c6}
  \Gamma[\bar g_{\mu\nu},\bar\lambda^{\mu\nu}] &=& -i\ln\Bigg\{
  \sum_{i} \exp i \int d^4 x \sqrt{-\left(\bar g+\kappa h^{(i)}\right)}
  \left(
      {\cal L}_g\left(\bar g_{\mu\nu} + \kappa h^{(i)}_{\mu\nu}\right)
    +\bar\lambda^{\mu\nu}\frac{\delta}{\delta h_{\mu\nu}}
 {\cal L}_{g}\left(\bar g_{\mu\nu} + \kappa h^{(i)}_{\mu\nu}\right)
\right.   \nonumber \\ && \left.
 +{\cal L}_{gf}\left(\bar g_{\mu\nu} + \kappa h^{(i)}_{\mu\nu}\right)
 \right)
{\det}^{-1}\left[\frac{\delta^2}{\delta h_{\mu\nu} \delta h_{\lambda\sigma}}  
\sqrt{-\left(\bar g+\kappa h^{(i)}\right)}\left(
{\cal L}_{g}\left(\bar g_{\mu\nu} + \kappa h^{(i)}_{\mu\nu}\right)+
{\cal L}_{gf}\left(\bar g_{\mu\nu} + \kappa h^{(i)}_{\mu\nu}\right)\right)\right]
\nonumber \\ &&
{\det}^2\left(\bar D^\lambda \bar D_\lambda \bar g_{\mu\nu} - \bar R_{\mu\nu}\right)
\Bigg\}  
\end{eqnarray}
In Eq. \eqref{c6}, the summation over $i$ is over all configurations of $h_{\mu\nu}$ that satisfy the classical equation of motion
\be\label{c7}
\frac{\delta}{\delta h_{\mu\nu}}\Bigg[
  -\frac{2}{\kappa}\sqrt{-\left(\bar g+\kappa h\right)} R(\bar g_{\mu\nu} + \kappa h_{\mu\nu}) +
\frac{1}{\xi}\sqrt{-\left(\bar g+\kappa h\right)}
\left(\bar D^\mu h_{\mu\nu} -\frac{1}{2} \bar D_\nu h^\alpha_\alpha\right)
\left(\bar D_\lambda h^{\nu\lambda} -\frac{1}{2} \bar D^\nu h^\beta_\beta\right) 
\Bigg]= 0.
\ee

If we were to set $h^{(i)}_{\mu\nu}$ in Eq. \eqref{c6} equal to zero, we would have the contributions of all diagrams to $\Gamma[\bar g_{\mu\nu}]$; these consist of all tree level diagrams (the exponential in Eq. \eqref{c6}), twice the usual one-loop diagrams involving the graviton (the contribution of the factor ${\det}^{-1}$ in Eq. \eqref{c6}), and twice the usual ghost-loop diagrams (the factor ${\det}^2$ in Eq. \eqref{c6}). No diagrams beyond one-loop order contribute to $\Gamma$. This expression for $\Gamma$ is consistent with unitarity \cite{McKeon:2021qhv}.

The functional determinants in Eq. \eqref{c6} result in divergences \cite{F4}; these are twice what appears in Eq. \eqref{eq1}. This divergent contribution to $\Gamma$ 
\be\label{c4N}
   {\cal L}^{div}=\frac{\sqrt{-\bar g} \mu^{4-d}}{4\pi^2(4-d)}\left\{
   \left[
\frac{1}{120} + \frac{1}{6}(\xi-1)^2
     \right] \bar R \,\bar g^{\mu\nu}
   +\left[
\frac{7}{20} + \frac{\xi(\xi-1)}{3}     \right]\bar R^{\mu\nu} 
   \right\} \bar R_{\mu\nu}     
\ee
($\mu$ is a renormalization mass scale parameter)
can be absorbed into
\begin{eqnarray}\label{c5N}
  \left.\bar\lambda^{\mu\nu} \frac{\delta}{\delta h^{\mu\nu}} \sqrt{-\left(\bar g+\kappa h\right)}
{\cal L}_g\left(\bar g_{\mu\nu} + \kappa h_{\mu\nu}\right)
  \right|_{h_{\mu\nu}=0} = 
-\frac{2}{\kappa^2}\sqrt{-\bar g} \bar\lambda^{\mu\nu}\left[
  \frac{1}{2}\left(\delta^\alpha_\mu\delta^\beta_\nu + \delta^\alpha_\nu\delta^\beta_\mu -
\bar g^{\alpha\beta}  \bar g_{\mu\nu}   \right)
\right] \bar R_{\alpha\beta}
\end{eqnarray}
by setting
\begin{eqnarray}\label{c6N}
\bar\lambda^{\mu\nu}_R &=& \bar\lambda^{\mu\nu} -
\frac{\kappa^2 \mu^{4-d}}{4\pi^2(4-d)}\left\{
   \left[
\frac{1}{120} + \frac{1}{6}(\xi-1)^2
     \right] \bar R \,\bar g^{\alpha\beta}
   +\left[
\frac{7}{20} + \frac{\xi(\xi-1)}{3}     \right]\bar R^{\alpha\beta} 
   \right\}
\nonumber \\ && \;\;\;\;\;\;\;\;\;\;\;\;\;\;\;\;\;\;\;\;\;\;\;\;\;\;\;\;\;\;
   \left\{
 \frac{1}{2}\left(\delta_\alpha^\mu\delta_\beta^\nu + \delta_\alpha^\nu\delta_\beta^\mu -
\frac{2}{d-2}\bar g_{\alpha\beta}  \bar g^{\mu\nu}   \right)
   \right\} .
\end{eqnarray}
All divergences are absorbed by the renormalized LM field $\bar\lambda^{\mu\nu}_R$. Neither $\kappa^2$ or $\bar g_{\mu\nu}$ are renormalized, and consequently they do not vary as the renormalization mass scale is changed.

Is there is a scalar field $\phi$ in addition to the metric field $g_{\mu\nu}$, then the classical Lagrangian of Eq. \eqref{eq2} is supplemented by
\be\label{24n}
 {\cal L}_{\phi}(\phi) = \sqrt{-\bar g}\left(
\frac{1}{2} g^{\mu\nu} \partial_\mu \phi \partial_\nu \phi - \frac{m^2}{2}\phi^2 - \frac{G}{4!}\phi^4
\right).
\ee
Quantizing $\phi$ in the same manner as $g_{\mu\nu}$ results in
\begin{eqnarray}\label{25n}
  \Gamma\left[\bar g_{\mu\nu}, \bar\lambda^{\mu\nu},\bar\phi\right] &=& -i \ln\Bigg\{
  \sum_{i} \int {\cal D} \psi \exp i \int d^4 x\left(
\sqrt{-\bar g+\kappa \hbar^{(i)}} {\cal L}_{\phi}\left(\bar\phi+\psi,\bar g_{\mu\nu}+\kappa h^{(i)}_{\mu\nu}\right) + {\cal L}_g\left(\bar g_{\mu\nu}+\kappa h^{(i)}_{\mu\nu}\right)\right)
\nonumber \\ &+&
\bar\lambda^{\mu\nu}\frac{\delta}{\delta h_{\mu\nu}}
           {\cal L}_g\left(\bar g_{\mu\nu}+\kappa h^{(i)}_{\mu\nu}\right) +
           {\cal L}_{gf}\left(\bar g_{\mu\nu}+\kappa h^{(i)}_{\mu\nu}\right)
           \nonumber \\ &&
                     {\det}^{-1}\left[\frac{\delta^2}{\delta h_{\mu\nu} \delta_{\lambda\sigma}}
\sqrt{-\bar g+\kappa \hbar^{(i)}} \left({\cal L}_{g}\left(\bar g_{\mu\nu}+\kappa h^{(i)}_{\mu\nu}\right) 
{\cal L}_{gf}\left(\bar g_{\mu\nu}+\kappa h^{(i)}_{\mu\nu}\right)\right) 
\right]
                     \nonumber \\ &&
                               {\det}^2\left(\bar D^\lambda \bar D_\lambda \bar g_{\mu\nu} - \bar R_{\mu\nu}\right)
           \Bigg\}  ,
\end{eqnarray}
where $\phi$ has a background component $\bar\phi$. The integration over $\psi$ leads to divergences that can be removed by renormalizing $m^2$, $G$ and $\bar\phi$, as well as further divergences proportional to $\bar R_{\mu\nu}$, which can also be removed by being absorbed into $\bar\lambda^{\mu\nu}$.
All divergences in Eq. \eqref{25n} are consequently eliminated.



\end{document}